# Value-based Engineering for Ethics by Design

Sarah Spiekermann and Till Winkler *Members of IEEE*

**Abstract**— This article gives a methodological overview of Value-based Engineering for ethics by design. It discusses key challenges and measures involved in eliciting, conceptualizing, prioritizing and respecting values in system design. Thereby it draws from software engineering, value sensitive design, design thinking and participatory design as well as from philosophical sources, especially material value-ethics. The article recognizes timely challenges for Value-based Engineering, such as the integration of ethical with functional system requirements, responsibility in hardly controllable ecosystems of interconnected services, active integration of external stakeholders and the difficulty in measuring the ethicality of a system. Finally, the Value-based Engineering methodology presented here benefits from shared experiences collected in the IEEE P7000 standardization process as well as from a case study. P7000[1] was set up by IEEE to establish a process model which addresses ethical considerations throughout the various stages of system initiation, analysis and design.

---

## 1 INTRODUCTION

THE past 10 years have confronted the engineering world with severe challenges related to the ethics of technology design. The Snowden revelations, a massive increase in cyber attacks, an explosion of hate speech, online addiction and the rise of surveillance capitalism [1] bring values such as privacy, security, freedom of speech and mental health—that have long languished in limbo—to the forefront of debate in professional IT system design. Moreover, the Boeing 737 Max crashes as well as the Volkswagen affair reminded the most prestigious engineering corporations how the values of long-term system sustainability, safety, transparency and human control can negatively interact with profit and speed of production. Against this background, the engineering world confronts an increased call for more "ethical" considerations during system design, a call that is typically equated with more value principles becoming part of an IT system's roadmap [2].

Various research communities have worked on values in computing [3], [4]. Scholars in the field of value sensitive design (VSD) [4], [5], machine ethics [6], [7], human-computer interaction, and participatory design [8] are known for their contributions. A wealth of research exists on how to respect privacy, security and safety in system design and, in particular, how to use risk-assessment based design to address these [9]. Agreement exists among most scholars of the field that an early recognition and tackling of stakeholder values in system analysis, design and development can ensure less value harms; harms that are diverse, context- and technology- specific, often not regu-lated, and therefore unexpected, unless an "ethics by design" effort has been made.

However, industry has not yet embraced ethics by design as a common practice. One reason for this is that 20th century economics has reduced the meaning of the term "value" to "profit". Only slowly—and potentially fostered by the more general sustainability movement—criticism is being voiced against this monetary understanding of value. The original significance of the term, denoting that "value" stands for worthiness, goodness or importance in its own right (regardless of money), has been revived. Business gurus call for "wise leadership" [10] and "shared value" creation [11], emphasizing corporations' social responsibility beyond profit and encouraging leaders to strengthen the value qualities and thereby ethicality of their products.

"Value-based Engineering" is a corporate innovation and engineering practice that caters to this new value-centered thinking. It supports a new era, where IT systems—the value bearers—are built to contribute to society's flourishing, while prohibiting negative side effects. The implication is that systems are not created any more merely because they maximize profit, are somehow useful, or embed new technical functionality; instead, technology's outspoken role is to support what is good, true, beautiful, peaceful and worthy in life. With this mission, Value-based Engineering seeks to extend non-functional requirements engineering, usability- or UX efforts. It is also different

---

- *S.Spiekermann is the chair of the Institute for IS & Society at Vienna University of Economics & Business. E-mail: wi-sek@ wu.ac.at*
- *T. Winkler is a Ph.D. candidate at the Institute for IS & Society at Vienna University of Economics & Business. E-mail: wi-sek@ wu.ac.at* from technology impact assessment approaches in that it does not focus exclusively on the risks of value harms. In contrast, it strives to build "technology for humanity."

[1] The IEEE P7000 Working Group (WG) has more than 30 experts with diverse backgrounds, including software and system engineers, but also in philosophy. The baseline for these experts' discussions was a P7000 draft that was written by the lead author of this article. This article benefits from learning gained from expert conversations in the group as well as more than 1,000 written impulses, critiques, and change requests shared in the standardization work group and collected in four P7000 commenting periods in 2017, 2018, 2019 and 2020. That said, this article solely represents the views of the authors and does not necessarily represent a position of either the IEEE P7000 Working Group, IEEE or the IEEE Standards Association.

The challenge of Value-based Engineering is getting industry to embrace it. Many companies today—and particularly those in high-tech—work in a process-driven manner. They need to know the activities and roles required to achieve ethics by design. When value creation is the goal––which is our understanding of ethics here—then companies need to know how to integrate values with the functional goals for their system design as well as with the ecosystem of partners they work with. For this reason, Value-based Engineering offers an integrated framework of concrete processes, activities and roles as well as a well-grounded terminology that can be followed. Because it provides such concrete processes, Value-based Engineering does not prescribe any particular list of value principles. Instead, it guides companies to discover for themselves what is relevant for them in their specific technology context.

In this paper we outline the processes of Value-based Engineering and show how they connect with prior literature. We provide definitions for all the main Value-based Engineering constructs, and derive 14 methodological requirements as well as 20 recommendations for companies who seek to apply it. As a proof of concept, we illustrate many arguments with a real-world case study we conducted with a Telemedicine start-up (hereafter referred to as "TM"). TM is a medical video chat-based diagnosis and recommendation service (see: [12]). Its unique selling proposition is the recommendation of highly ranked medical specialists. The TM case is one of four case-studies that we conducted to test Value-based Engineering. Furthermore, Value-based Engineering has evolved in recent years with the help of system engineering experts who contributed to a five-year-long IEEE standardization project, conducted under the acronym "P7000." P7000 was set up by IEEE's Computer Society in 2016 to establish a process model for addressing ethical considerations throughout system initiation, analysis and design. The lead author of this article was vice-chair of this standardization project, main author and editor of the initial drafts as well as a co-initiator of its setup. Whenever reference is made to "experts" in this paper then these refer to members of the standardization work.

When hereafter using the term "engineer(s)" it is important to note that we do not only refer to system and software developers. Rather, we recognize that a system is nowadays often co-determined by larger project teams that include computer scientists, electrical engineers, product managers, legal scholars and many more. In close cooperation these individuals "engineer" a system that they want to bring to market. So, engineers are broadly defined here as those members of a team who influence the values, goals, architecture, data flows, policies, hardware and software components of a system.

Finally, we want to stress that we integrate to the best of our knowledge other scholarly efforts on values in system design as well as non-functional requirements engineering. We want to especially acknowledge the VSD community, which we consider as the main predecessor of Value-based Engineering. Our framework and methodology presented here shares many roots with VSD that we acknowledge in detail throughout. Value-based Engineering extends VSD in three crucial points: First, we complement the value elicitation phase of VSD by proposing how concrete ethical theories can guide the envisioning of relevant system values. The value ontology embedded in Value-based Engineering is grounded in centuries of philosophical ethical reasoning and has a nuanced terminology stemming from the philosophical field of material value-ethics. Second, we detail the engineering part of innovation: what VSD calls "technical investigation" [13]. We show precisely how value principles can be translated into system requirements. Third, as outlined above, we want to go further than VSD, encouraging companies to reconfigure their entire innovation processes and "base" these on human values as their bedrock.

## 2 WHAT ARE VALUES AND WHY SHOULD ETHICS BY DESIGN FOCUS ON THEM?

The idea of focusing on "human values" when designing technology could not be taken for granted when VSD scholars first introduced it in the 1990s [4]. Human values will always remain contextually malleable phenomena accessible more to our "emotional intuition" [14, p. 272] than to any objective measures. They are less accessible than norms, that are repeatedly observable, or rights, which are articulated in the law. So why base engineering effort on them specifically? We would argue that the value concept is in essence more inclusive and wider than human rights, social norms or needs. Human rights naturally correspond to what is valuable in humans' lives [15, p.64 ff.]. Yet there are more human values at stake when building technologies than there are established human rights. For instance, values such as beauty, human control, or transparency are not established in human rights. For this reason it is wise to focus on the wider concept.

Besides human rights, some computer ethicists have focused on *norms* as the driving force of ethical system design [16]. Norms are culture-specific representations of acceptable group conduct. Norms can be problematic though when abused by normative powers—for example, when politicians establish total transparency as a norm that allows them to survey citizens. Therefore, the positive value(s) that a norm fosters in society should first be challenged in a non-hegemonistic way. Positive values are a pre-condition for identifying good norms for system design.

Along with human rights and norms, some scholars have proposed a focus on *needs* in order to build ethical systems [17]. The popular "Design Thinking" schools have been using needs as their conceptual baseline for innovation, going far beyond functionality-driven forms of innovation. However, many negative values of today's IT systems, such as a loss of privacy and freedom or a lack of transparency, are not recognized as a need by many end-users. While these are important values from an expert perspective when it comes to ethical engineering, direct stakeholders have difficulties in grasping these abstract concepts in emotional terms as something they need. In contrast, positive values allow us to embrace the good, true and beautiful in their own right. Thereby they trigger a kind of motivation in people that goes beyond what might be immediately necessary. Taken together, values are again the broader concept: they include needs, because only what is valued is needed.

Against this background, it does not seem to be an arbitrary development that the most important pioneering groundwork for ethics by design has been done in the past 20 years by the VSD community [4], complemented by other scholars who adopted the concept of values [3], [18], [19]. Through many conceptual and applied research efforts, the concept of values has become the core of many established university textbooks on computer ethics [6], [18] and well-regarded technology design books [20], as well as the center of political calls for engineering ethics [21]. Finally, it is interesting to observe that also among management scholars [10] and economists [22], this focus on "values" can be observed, specifically in the form of virtue ethics.

But what are values? Are they equal to virtues? Values are defined as "a conception … of the desirable which influences the selection from available modes, means and ends of action" [23, p. 395]. They are "principles of the ought-to-be" [15, p. 98]. That is, however, only when they are positive, since according to material value-ethics, values can also be negative [24]. For example, the value of privacy is the positive pole of a value spectrum where the breach of privacy would be seen as a negative value. Material values are not just subjective preferences, opinions or cultural conventions; rather, like geometric principles, values are given a priori [15, p. 135f], similar to what Plato called "ideas." A company CEO might hold the opinion that privacy is unimportant or outdated, but he or she cannot change the fact that the value of privacy exists. It exists independently of the CEO's personal or cultural preference and can therefore be appreciated by customers. This appreciation is given as a truly felt attraction; or, in the case of a negative value, a truly felt repulsion. "We become aware of values in acts of "feeling," writes Kelly [14]. Such acts of value feeling can also come into play when we observe the behavior of others. When we observe friends becoming hateful, impatient or addicted to social networks, we see that the value of their human conduct—also called "virtue"— has suffered.

## 3 PREPARING FOR VALUE-BASED ENGINEERING

Value-based Engineering can be implemented when a new technology is first created (greenfield situation) or it can be applied to an existing service (brownfield situation). In a greenfield project, the starting point is an unconstrained initial service idea combined with a rudimentary concept of operation. Such a concept of operation is a "verbal and/or graphic statement, in broad outline, of an organization's assumptions or intent in regard to an operation or series of operations" [25, p.4]. It roughly sketches out the envisioned "system of interest" (SOI), with its core elements, stakeholders, data flows, interfacing systems and context-of-use descriptions. The TM case study we refer to hereafter (for more details see [12]). TM had the vision that patients dial into its platform to speak via video with a TM general practitioner and get a first diagnosis. Its doctors give initial advice, and write prescriptions and sick notes; but most importantly their task resides in passing a patient on to the right specialist for their need. For example, if a kidney problem is diagnosed, the TM doctor could access TM's database of kidney specialists in the patient's respective region and recommend not just any kidney specialist, but one particularly well rated by doctoral peers.

Note that when we speak of a "system" or "system of interest" (SOI) like in the TM case we always refer to a socio-technical system. Socio-technical systems regard technology as embedded in organizational, public or private processes, which include policies, people, preferences and incentive systems [26].

### 3.1 Recognizing ethically relevant system boundaries

When seeking to anticipate the ethical effects of an SOI, its technical boundaries need to be well understood. Most companies today operate within a wider "system-of-systems" (SOS). At their organizational boundaries they interface with external web-services, databases and code components (third-party systems). TM for instance planned to rely on a video chat application from an external video service provider. It wanted to store patients' health data in a remote cloud-service. So at least two external partner-systems were foreseen to be integrated into the SOI structure that would present itself as one system to users in the end. An SOS can influence the values created for end users. If, for example, the cloud provider of TM does not handle the health data it stores in a secure way and gets hacked, patients will consider TM as untrustworthy no matter how well TM's own systems are secured. **SOI operators that pursue Value-based Engineering therefore need to embrace responsibility for the values and operations of their SOS ecosystem (Requirement 1).** While such a "principle of responsibility" has been recognized philosophically it has practical limits: Official managerial influence on adjunct systems is limited by system ownership. Consequently, ethical service delivery depends on the cooperation of a whole network of "well behaving" partners; otherwise, the whole SOS can be negatively impacted by one "black sheep." **Value-based Engineering anticipates this possibility and therefore recommends that organizations include at least all first-tier SOS partners in the ethical analysis as well as those that bear a chance of ethical import (Recommendation 1).** In particular, those SOS entities should be included in the ethical analysis that process or provide any personal data relevant for the SOI.

Such an extended responsibility is of course only feasible if SOI operators have sufficient insight and influence on what partners are doing. Value-based Engineering therefore recommends a governance and control-analysis of all the elements of an SOI's SOS in line with ISO/IEC/IEEE 15288. This standard distinguishes between virtual, collaborative, acknowledged and directed types of SOS [25]. In acknowledged forms of cooperation, the constituent system owners retain independent ownership, management and resources, but they recognize joint objectives (i.e., value objectives) and designate people responsible for managing these objectives. Directed forms of cooperation go even further: Here, an SOS is built and managed as one entity. The component systems maintain their ability to operate independently, but the normal operational mode is subordinated to the joint and central purpose. **In both of these forms of partnership the recommendation to test for controllability is that the SOI operator can obtain access to the enabling system elements of the SOS (Recommendation 2).** In TM's case this was a challenge. Cloud services are for instance typical commercial off-the-shelf services that allows for little influence on their operations. TM therefore needed to seek out a particularly transparent and open cloud partner to offer patients the privacy that a health application requires.

### 3.2 Involving stakeholders openly and extensively

Once the broad operational concept and the boundaries of an SOI are clear, it is possible to complete a list of stakeholders, that is, those "individuals, organizations, groups, or other entities that can affect, be affected or perceive themselves to be affected by it. Stakeholders have a legitimate right, share, claim, influence or interest in a system …" [27]. Two stakeholder categories should be discerned: First, those who directly interact with a technology, such as human beings directly using the system (end users) or organizations purchasing the system (acquirers). And second, those "stakeholders who, although they never or rarely interact with the system as end users, are nevertheless affected by the system" [4, p.38]. These are called indirect stakeholders [4]. Examples are communities, neighborhoods, institutions, nation states, future generations, but also animals, nature or entities with historic or sacred meaning [4].

Since an SOI normally affects many different stakeholders, only an extensive and diverse stakeholder involvement has a chance to anticipate a reliably complete value spectrum for the SOI. Therefore, experts in VSD [4] have been arguing that ethical design efforts should work with stakeholder representatives who sensitize for minorities (such as NGOs), are critical of the SOI, and represent the interests of target market users. In line with this established view [4], [8] **Value-based Engineering requires that an SOI must be designed in cooperation with an extended group of representatives of both direct and indirect stakeholders (Requirement 2). Furthermore, it is recommended that any international rollout of technology should be accompanied by the inclusion of stakeholder representatives stemming from those regions of the world in which a system will be deployed (Recommendation 3).** In the case of TM 20 stakeholders were named. Direct stakeholders included students who might seek sick leave on exam days, elderly people who would benefit from the service but don't have internet access, and foreign minorities who don't speak the national language and are not insured. Indirect stakeholders included young doctors whose reputation was as yet insufficient to garner a good rating, and the doctoral community as such.

Ulrich [28] showed that the sources of stakeholder motivation, power and knowledge, as well as their legitimation, should be considered in their selection. Once the right stakeholder representatives are on board it must be ensured that they can effectively bring in their voice. This is not a given, because organizational hierarchies, hidden agendas or other social dynamics can impede an open exchange of critical perspectives. Mingers and Walsham [29] have therefore pointed to traits of ideal speech situations [30] to ensure that such dynamics do not undermine the discourse. Stakeholders should be allowed equal participation, be encouraged to question claims and assertions and be able to freely express attitudes, desires and needs [18].

### 3.3 Envisioning, documenting and monitoring the relevant context

A challenge for Value-based Engineering is that ethical behavior is always deeply contextual. Even though there are philosophies such as duty-ethics, which promote the use of universal value principles regardless of context, history has shown that this is problematic. Take the example of the duty to always speak the truth. Now imagine a humanoid household robot programmed simply with this duty-ethical rule [31]. Furthermore, imagine its deployment in a household hiding a political dissident who rightfully fights an oppressive political regime. When opening the door to the regime's police officers, the household robot would tell the truth and reveal the hiding place of the dissident that it should have concealed. Despite its duty-ethical algorithm, it might behave in an ethically questionable way unless it was sophisticated enough to understand the socio-political context. The example shows that the context of a computer system's future deployment must be deeply understood and anticipated before developing and deploying it. Noteworthy context elements in this example are the political surroundings in which the system operates, the long-term horizon assumed and the infrastructure available. These elements are not typically considered today in context-sensitive computing efforts. Commonly used context factors include a user's location and environment, identities of nearby people and objects as well as changes to those entities [32]. A standard like ISO/IEC 25063 [33] includes a helpful common industry format for describing a context of use. However, it is only focused on the usability and not the ethicality of an SOI. It looks at how user dispositions influence an SOI, but not the other way round; that is, how an SOI might influence user dispositions, which is the more relevant dynamic in Value-based Engineering. A requirement for the preparation of Value-based Engineering is therefore to go beyond current context-modeling approaches. **Innovation teams need to physically explore the context(s) of SOI use where possible, or envision what this context would look like if the SOI was deployed at scale (had big market share) (Requirement 3).** Envisioning can be done with the help of creativity techniques as have been suggested and trialed by the VSD community [4]. Likely contexts should then be described in detail before the ethical analysis of system deployment can start for each of them.

What should such a contextual description contain? Respecting Helen Nissenbaum's call for guarding the "contextual integrity" [19] of data we recommend that a context description should include the data processing elements of the SOI (including relevant 1st tier SOS elements), the data flows between them, the data subjects involved, the data types processed and any ethically relevant aspects of integrity linked to this setup. For TM we can for instance distinguish data subjects (patients, specialists, recommending doctors and TM doctors) and a data recipient (TM). All are subject to various data types. One type is patients' health information. Integrity expectations might be that this data is collected with consent, and treated confidentially and with no commercial interests attached. The other type of data is ranking data obtained from TM's recommendation community (data senders). Integrity expectations might be that this data is also consent-based and confidential and non-commercial, at least from the perspective of the specialists who are ranked. TM, however, might have a different perspective since it potentially wants to exploit these rankings commercially. The example shows that ethical questions already arise at the context-modeling stage, which is why **we recommend considering this contextual description task as an opportunity to grasp a first set of potential ethical issues around the system (Recommendation 4).**

Envisioning all the involved entities in a concept of operation, recognizing different use contexts, considering long-term pervasiveness and then nailing this down into a document is time-intensive. In addition, the true future contexts of systems can only be anticipated to a limited degree. Only the real-world deployments and uses as well as abuses of an SOI will reveal the full spectrum of what might be ethically relevant. **Organizations should therefore capture the context of system-use not only prior to system design, but also after deployment (Recommendation 5).** This monitoring requirement for context evolution is consistent with the iterative nature of the VSD approach.

Finally, when asked to envision deployment context(s), some experts argue that their systems are initially of such a generic nature that the context(s) of their later use is not known. For example, when computer vision algorithms are developed that translate a pixel space observed into a precise picture representation, then this kind of technology could be used in so many contexts that it seems unfeasible to describe these in the way illustrated here. Computer vision may be deployed in contexts ranging from cancer recognition applications to military drone targeting systems. However, Google's "Project Maven" case demonstrated how engineers were unpleasantly surprised when their generic computer vision technology was suddenly foreseen for military purposes [34]. **In light of this case, Value-based Engineering recommends considering context-of-use scenarios as early as possible, and at the latest when a system is trained for or applied in a known industry (Recommendation 6).** There is a point relatively early in system design where a generic system is adapted to serve its final use. This is the point, for example, where the computer vision algorithm is applied to and trained with data from either a military or a health context. It is recommended to begin with context analysis of the kind recommended where a technology is thus applied to a concrete use-context.

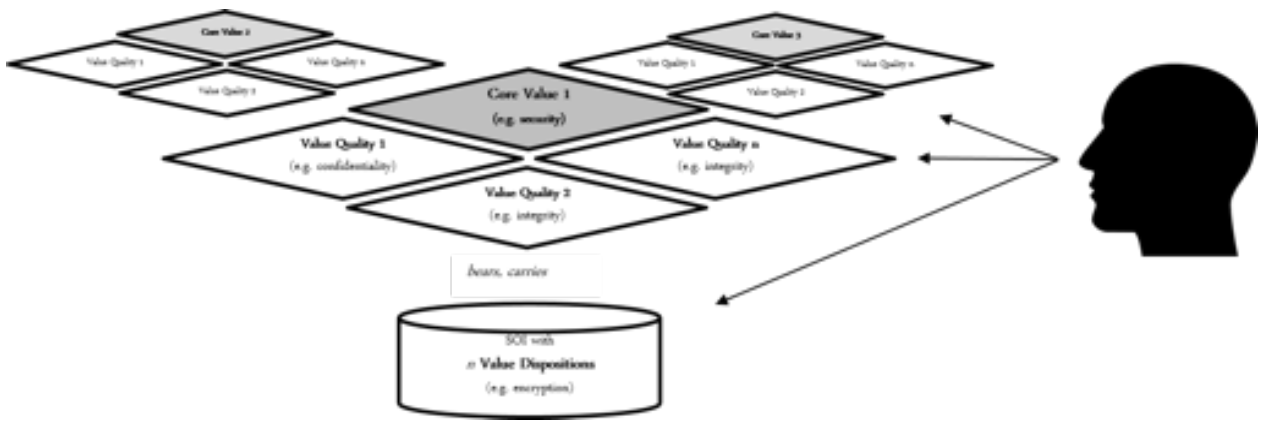

Fig. 1. Value space phenomena and terminology

## 3.4 Adopting a consistent value terminology

Using values for system engineering requires that we understand precisely what they are. A challenge for using values in system design is that every technology comes with a large number of potentially relevant value consequences. In the TM case, for example, we identified 214 negative value potentials and 253 positive ones. Such a large number of values was confirmed by three other case studies. Delving deeper into positive values, we found 108 of them being *instrumental* in nature, such as the desire of users to control their health data. Such data control caters to the higher and *intrinsic* value of privacy. Scholars working on values in system design have recognized the complexity. Kelly writes, "…values condition each other, in that it is not possible to grasp one value without having grasped some others" [14, p. 11]. Unfortunately, no clear ontology of terms has yet been presented. Scholars talk fuzzily about various "levels of [value] abstraction" [35], [5] to describe value relationships or hierarchies [24]. Somewhat heretically, an analogy could be drawn with the early days of electrical engineering: "voltage," "ampere" and "watt" were at some point noted to be different things. Just talking about voltage—or analogously, values—does not suffice to properly acquire the domain for designing a system.

In this vein a contribution of Value-based Engineering is that it suggests an ontologically differentiated terminology to describe the value space, which is philosophically derived (at least in part) from Max Scheler's "Formalism in Ethics and Non-Formal Ethics of Values" [24]. In 1921, Scheler discerned various elements of the value space that can be useful when referring to system design. In particular he described "value qualities" catering to what we call "core values". He discerned "value dispositions" from "value bearers." Without knowing about this, engineers are already used to discerning core values and value qualities. For instance, when we say that the core value of "security" is achieved by the "confidentiality," "integrity" and "availability" of a service. The core value (security) shows itself in a multitude of value qualities (confidentiality, integrity, availability. Value dispositions, such as encryption, ensure that the SOI can be perceived as secure. And the SOI is then called a value bearer, because it bears the core value of security (see Figure 1).

Applied to TM its CEO argued that no matter what financial means or insurance status, every patient should have equal access to good healthcare specialists, a privilege that is often reserved for the richer people in a society. Therefore, equality was a core value to him. Four value qualities were identified to realize this vision: TM's service would need to be inclusive, allowing poor and uninsured people to use it (value quality 1: inclusion). Exclusion (value quality 2) of those who do not have a computer or are anxious about using it should be avoided. True equality would furthermore only be achieved if the specialists recommended by TM's database were also willing to treat the patients flowing in from TM (value quality 3: access). And finally, the virtual encounter between a TM doctor and a patient should not be accompanied by any kind of biased information that could lead to a loss of care on the side of TM (value quality 4: loss of care). Against the background of this example it becomes clear why value qualities have been described as the "real qualities of [core] value itself"

[24, p. 6] by Max Scheler. Experts engaged in the IEEE P7000 project still opted to style value qualities rather as "value demonstrators" and defined them as the "manifestations of a core value, which are either instrumental to a

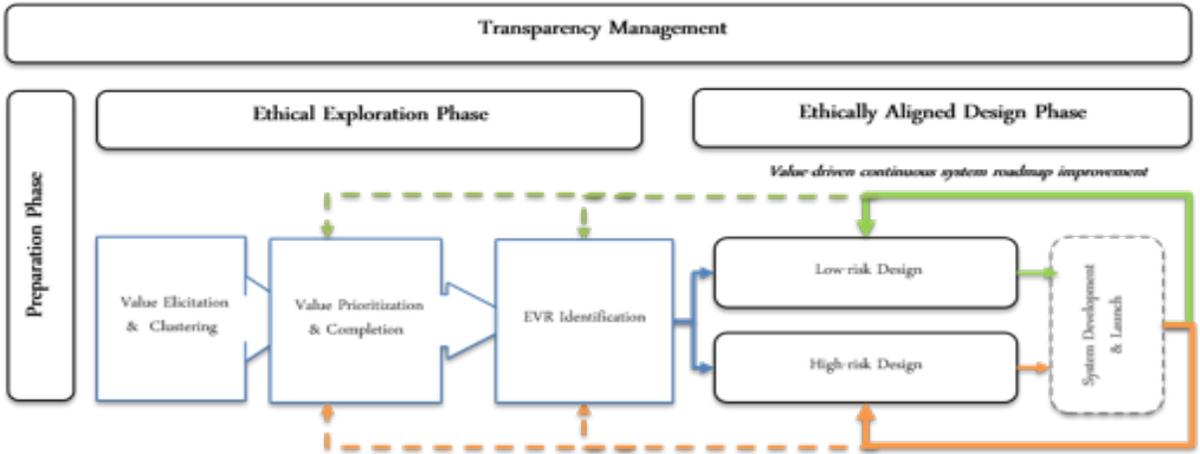

Fig. 2. Overview of Phases and Processes recommended for Value-based Engineering

core value or undermine it" [27].

The example illustrates that values and their qualities can have technical implications even when they are not necessarily of a technical nature. For instance, if bias against uninsured patients should be avoided, then one technical disposition for this might be that no insurance status should be displayed to TM doctors in TM's interface design. A value disposition "creates the potential for one or more [positive or negative] values to unfold when using or examining a system" [24, p.79]. It is a technical or organizational prerequisite created by engineers and embedded in the product or service architecture.

Value dispositions can be built into an SOI, while core values and value qualities cannot. Instead, value qualities––such as those we identified here for TM equality—can only be "experientially present 'on' the physical objects, acts, and persons we encounter…" [14, p. 19]. This means that computer systems do not "have" values, but they "bear" or "carry" them—if they have the necessary value dispositions built into them. If this is the case, human beings can feel attracted to the value qualities created by them. Figure 1 summarizes the fine distinction of terms, here once again taking up the example of security. **We strongly believe that embracing this distinct terminology facilitates communication about values. We therefore recommend using it when working with values in system design projects (Recommendation 7).**

## 4 HOW VALUE-BASED ENGINEERING WORKS

Value-based Engineering consists of the two phases of ethical exploration and ethically aligned design, each made of a number of sequential as well as iterative processes depicted in Figure 2. If one discussed Value-based Engineering in relation to the traditional Waterfall SDLC (system development life cycle), most of its novel contributions to system engineering would be found in the early phases of it. Hereafter, we therefore describe in greater detail how three processes of Value-based Engineering cater to this early SDLC phase: value elicitation, value prioritization and ethical value quality identification. These three processes constitute an "ethical exploration phase," where "ethical" means "value-based." Once a company has gone through ethical exploration, it then aims to ensure that those values it has prioritized and conceptualized are effectively finding their entry into the technical and organizational design of an SOI. The "Ethically Aligned Design phase" takes care of this. It includes two complementary processes, a low-risk iterative design process and a more time-consuming "risk assessment-based design process." Both of them foresee the systematic identification of system architecture and design treatments for relevant values at risk. They differ, however, in their levels of depth, documentation and stakeholder involvement.

## 4.1 Ethical exploration phase

The ethical exploration phase envisions relevant values associated with an SOI: it prioritizes and completes them, it shapes their role in the business mission, it anticipates ethical pitfalls and it decides on further SOI investment. **It is recommended that this phase of work is supported by a new kind of employee: a 'value-expert' who has knowledge on values (i.e., training in and experience of material value-ethics), a strong faculty of speech, an aptitude for conceptual thinking and system thinking as well as an understanding of the legal and technical "worlds" (Recommendation 8).**

### 4.1.1 Value elicitation

Envisioning of values is not new to the HCI community. The core value of usability for instance is related to joy, convenience, user control, etc. VSD has driven the effort further by embracing further stakeholder values beyond the scope of classical HCI. In doing so, VSD does not commit to a singular ethical theory that might be used to elicit values [4]. This openness to any theory of ethics for value identification comes with a price though: Many VSD practitioners tend to identify values by only thinking about stakeholder benefits and harms [4], thereby intuitively following a form of utilitarian ethics. Manders-Huits (2011) therefore recommended that value-oriented design processes should be guided by more specific ethical theories, which are chosen and applied conscientiously [36]. **Value-based Engineering does this by requiring the use of guiding questions derived from three grand ethical theories for value elicitation that are typically covered in computer ethics textbooks (e.g. [37]): utilitarianism, virtue ethics and duty ethics [18] (Requirement 5a):**

 1. What are all thinkable positive and negative consequences you can envision from the system's use for direct and indirect stakeholders? (utilitarianism)
 2. What are the negative implications of the system for the character and/or personality of direct and indirect stakeholders—that is, which virtue, harms or vices could result from widespread use? (virtue ethics)
 3. Which of the identified values and virtues would you consider as so important (in terms of your personal maxims) that you would want their protection to be recognized as a universal law? (duty ethics)

By using these three questions, we aim to ensure that the values envisioned and the system ideas collected are ethically guided. A critique of using these three questions might be that utilitarianism, virtue ethics and duty ethics are all part of the Western philosophical canon, whereas Value-based Engineering aims to be used globally. While this is true, virtue ethics—at the level of abstraction used here—is practiced in most cultural traditions [26]: A virtuous person, for instance, while called *phronimoi* in the Aristotelian tradition, is called a *junzi* in Confucian ethics. And the Buddhist concept of *śīla*, i.e., a character upholding proper conduct, is exactly the kind of character Value-based Engineering seeks to protect. Also, we would argue that every culture and religious tradition has duties it holds dear. "Personal maxims" refer to subjective laws or principles of behavior according to which a person thinks she should act; not because she is forced to, but because she believes in the duty towards a "good" society by respecting them. We expect that stakeholders from different cultural traditions will vary in what they believe to be a duty towards a good society and how systems can support it. They will point to different virtues and duties important to them, many of which might never have been considered so far in technology design and may not even be preferred by a Western mindset. **In addition, Value-based Engineering requires that this analysis is complemented by a fourth question that explicitly embraces a culture-specific philosophical framework from the region of the world in which the SOI is deployed (Requirement 5b).** Cultural traditions have their unique ways of framing ethics, and thereby shed additional light on what matters to stakeholders.

In the TM case we just used the three Western philosophical questions. At the outset its detailed business plan contained 12 "mainstream" values anticipated for the platform, including efficiency, convenience and flexibility for patients and doctors due to fewer physical encounters. TM expected a health improvement for patients from its service due to specialist recommendations as well as linked cost advantages for insurers. TM also recognized that health data would need appropriate security and privacy measures. Applying the three ethical questions to the company case unveiled a more complete situational value potential that should be relevant for an investor in the company: As mentioned already, 214 potential negative value effects could be identified by the stakeholder representatives along with 253 positive ones. In a nutshell, stakeholders only saw TM improve the health of patients where it was able to win the sustained support of the medical community, which needs to continue giving "objective" and honest specialist recommendations for TM's specialist database. This support is not to be taken for granted though. Mutual ranking and rating of colleagues can breed competition among doctors, greed and dishonesty, issues identi-fied particularly by the virtue-ethical analysis. Stakeholders saw that TM could also breed dishonesty among patients, who might abuse the service for quick referrals, medication or sick notes. Furthermore, TM could undermine health, because good diagnosis requires a human touch, and TM has to strike a fine balance between the digital shortcut and the analog virtue of doctor–patient encounters. The efficiency aspect of the telemedicine platform carries the challenge that trust and patience between doctors and patients is undermined, and today's respectful relationship may be supplanted by the commodification of a discipline. Privacy finally can not only be undermined by unexpected secondary data uses mitigable through transparent policies, but it can also be positively fostered by TM in that it allows shy patients to remotely access medical advice in delicate affairs. The highest number of value-related ideas were found by the utilitarian question; the most original ones on personal character and relationship changes by the virtue-ethical analysis [38]. How was this first realistic value impression used for system design?

### 4.1.2 Value clustering

When stakeholders describe the issues they see for an envisioned SOI (with the help of the ethical guidance described above) they are not always able to perfectly frame their thoughts. For example, one TM stakeholder representative wondered: "What if patients are abusing the video chat with doctors and lie about their true condition just to get a quick sick note? Doesn't TM's virtual encounter encourage such a lack of accountability?" Looking at the unstructured text of this remark, relevant values are not all obvious, but need to be teased out. Two values can be extracted: honesty and accountability. Accountability was directly named, while honesty was hidden in the description and needed to be named by the value expert. To ensure that value experts are able to tease out what was really meant, **Value-based Engineering requires stakeholder representatives to always name the values they care about when they describe an issue (Requirement 6).** However, the TM case showed that while some values keep recurring, they can appear under different names. For instance, privacy was reflected upon in terms of the degree of control patients might want over their health data, the security of health data, the idea of staying anonymous vis-à-vis TM doctors, etc. The value expert needed to recognize that personal data control, security and anonymity all function here as three distinct value qualities of that one underlying core value of patient privacy. This required substantial abstraction and structuring work on the side of the value expert. In fact, in the TM case the 467 positive and negative value potentials were retrieved. And their naming, their refinement and their grouping into core values and value qualities took several days of analysis. **In Value-based Engineering the value expert bears this prime responsibility of analyzing the collected material and of distilling core value clusters as depicted in Figure 1 (Recommendation 9).** In the TM case, 14 core value clusters were distilled. A core value is unique to the respective context of the SOI. It is defined as a value repeatedly described by stakeholder representatives as relevant and therefore judged as centrally important by the value expert [27]. One way to determine what is "core" (and what is not) is to count all values of the same name and to define those that are most frequently repeated as core. This was how the core values of TM were determined. We stress, however, that the most prevalent values found in this way can overshadow more essential ones that might be mentioned less often. Therefore, a value expert should always trust his or her own judgment of relevance in defining what is core. **That said, s(he) should then reconfirm with stakeholder representatives a) that the core values chosen are the right ones and don't overlook anything and b) confirm that the expected value quality effects correctly capture the issues raised (Recommendation 10).**

### 4.1.3 Value cluster prioritization and value-mission building

Once core value clusters are identified, engineering teams need to decide how these should be prioritized. **Value-based Engineering foresees the active involvement of corporate leaders in this prioritization (Requirement 7)**, especially in order to avoid later value conflicts among engineers during development [3]. Three complementary analyses are used to prioritize core values: The first investigates how core values resonate with the existing or emergent business mission. The second analysis is a duty-ethical one. And the third analysis requires organizations to check core values against existing corporate principles, legal frameworks, international human rights agreements or relevant ethical principle lists. For TM, the first business-oriented analysis showed that out of the 14 core values identified for its platform it would be well advised to either focus on patient equality or on doctors' knowledge creation. Depending on the core "value proposition" chosen, the other core values would also take different priorities.

TM's CEO decided to prioritize the value of equality. Examples of other corporations' principles are IBM's commitment to accountability, explainability and fairness of its products [39], and Microsoft's commitment to people empowerment, community and environmental sustainability [40]. Such self-imposed duty principles show their ethical worth when they are used to effectively guide system design's value priorities. When corporate leaders are involved in value prioritization, they should therefore query the personal maxims they hold (in line with duty ethics as described above). **Value-based Engineering requires them to engage in some introspection and support only those core values as future system principles that they would want to become universal and would therefore publicly endorse (Recommendation 11).** Duty ethics also asks them to not treat other people as "a means only" to achieve their goals. If for instance a service is created primarily to create profit from selling a customer's personal data, then customers are used as a means only to serve financial gain. Value-based Engineering would consider this an unethical prioritization. A "higher" value, such as customer joy, or customer knowledge must be prioritized for system design in order to create technology for humanity.

A challenge that can occur at this point is that values sometimes contradict each other and require engineers to make a trade-off in ranking one higher than the other. Value-tradeoffs are often seen as a dilemma. Material value ethics does not emphasize the conflict among values, but rather sees them being in a natural hierarchy [24]. It claims that ethical behavior is constituted by choosing and realizing higher values over lower ones [14], [40]. Scheler described how the relative endurance, depth and indivisible nature of values are criteria for their superiority; their relative independence from value-bearers and the degree of intrinsic value they have are also a sign of their "rank" [24, p. 86ff]. Two values from TM's case illustrate this: Doctors' collegial respect that shows itself in unbiased mutual recommendation may be traded in for the efficiency provided by a database. Efficiency, however, is a lower value than respect. Unlike respect, efficiency has little intrinsic worth. One can ask, "What is efficiency good for?" and debate that sometimes—for instance, in human relationships—efficiency is not intrinsically "good. " Respect in contrast would not be doubted as worthy in itself; it unquestionably leads to deeper satisfaction in people than efficiency does. So the great metaphysical value conflict TM confronts is that it may promote a value of lower priority if respect is subsumed to efficiency. This idea of a value hierarchy can refine the VSD conception of values, in which values are seen in a delicate (but not necessarily hierarchical) "balance" with each other. A so-called "value tension" is typically solved by balancing values against each other, keeping them intact and not excluding any value [4].

Finally, experts have suggested a number of provisions that can be made for good value priority choices: First, Value-based Engineering requires transparency. It requires binding the value priorities made to individual executives who have to openly endorse their decisions. Furthermore, experts recommended that the wide group of stakeholder representatives described above should by and large support the prioritization process. **The final decisions on value prioritizations should not be taken by top executives' "brute force," and not without the true consent of the stakeholder representatives (Recommendation 12). Feedback cycles from service stakeholders are required on the value priorities of the SOI once it is launched (Requirement 8).** Such feedback supports the continuous improvement of a system so that value priorities can mature as the system matures. Finally, experts recommended that companies do not fully decide the value list for themselves. They think that companies should also check target market regulation and international human rights agreements. Such **legally recognized ethical principles provide the outermost boundary condition for corporate action and should impede the prioritization of lesser values (Requirement 9).**

The reflection on duties, corporate and legal principles as well as value hierarchies should lead innovation teams and leaders in some cases to decide against investment in a new product or service. **This active consideration to not invest in a SOI on ethical grounds must be seen as a critical part of Value-based Engineering (Requirement 10).** Value-based Engineering is not only about value elicitation and prioritization, it is also about an imperative to re-nounce business opportunities and profits that disrupt society and/or human well-being.

When prioritizing core value clusters with TM's CEO it became clear that the ranking of the SOI's core values (and their respective value quality clusters) would deeply influence the business mission. We spread out all 14 value clusters on a table and discussed each of them with the CEO, shifting and moving around the cluster images, discussing TM's "value-mission" resulting from different core value prioritizations. Following this discussion, we recommended that TM put the core value of knowledge first, framing TM's mission as a platform for peer-to-peer doctor support. Specialists and TM's general practitioners would then virtually support each other's diagnoses in difficult cases and thereby foster cooperation and lifelong learning in the community. Such a usage of the platform would have supported the long-term support and respect TM needs in the medical community and improved the health service in rural areas. TM's aspired-for database with mutual referrals would grow as a natural result of an online community of specialists and doctors supporting each other for the health benefit of the patient. Reliability and accuracy of health information would be highly important in such a value-strategy, but with an emphasis of making this information accessible between doctors. Given this sharing of health-related information between doctors, patient privacy would be among the top five value issue to tackle. The whole platform would need to be built towards being a highly secure, high-quality cooperation platform for experts, which aims to store and share patient data for a higher reason. In contrast to this potential value mission, TM's CEO chose the equality strategy briefly outlined above. Here TM's patient interaction and data storage would be kept to a minimum, which reduces some cost for privacy. Patients would be passed on quickly to the right specialist and without any further interaction. TM would be more of a help desk to everyone with a lean infrastructure. That said, this help desk would need to be trustworthy in the long-run, and accurate, the second and third highest core values in this value mission. Somehow TM would need to ensure that its specialist database sticks to what it promises; that is, high-quality referrals. We recognized that the exclusion of some doctors in TM's specialist database could lead to distrust and that a help desk vision treating everyone equally, but no-one rigorously, could lead to dishonesty and a lack of reliability on the side of patients who might abuse TM for quick referrals, sick leaves, etc. The reason that we describe the two strategies in such a detailed manner here is to show that the way in which leaders, stakeholders and engineers prioritize core value clusters can result in a very different service vision, that in turn is likely to imply a need for a very different system architecture and technical priorities. Here it becomes clear that Value-based Engineering is not only a technical endeavor, but a holistic strategic effort that integrates technology and management.

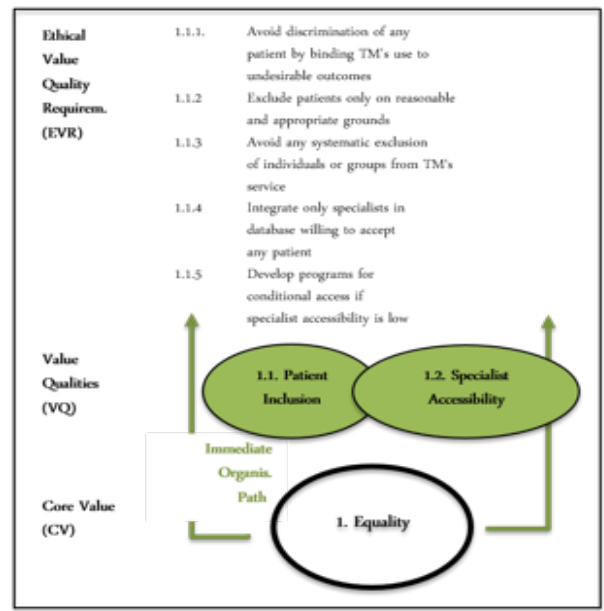

Fig. 3. The Value Register contains a chain from core value to value quality to EVR

Based on the value prioritization chosen, Value-based Engineering recommends the formulation of a public ethical mission statement. **This value mission statement summarizes the value priorities committed to in a nutshell, and is signed by organizational leaders (Recommendation 13).** For TM this could read, "The company's core goal is to create a health recommendation help desk where any patient—regardless of insurance status or money—can find online advice on health problems and be reliably referred to the best regional specialists. TM will constantly strive to improve the trustworthiness and accuracy of its recommendations and cares about the privacy of its patients."

### 4.1.4 Ethical value quality requirement (EVR)

After value elicitation, **Value-based Engineering requires a conceptual investigation (Requirement 11).** This is a hermeneutical exercise also recognized as a crucial step in VSD [4]. "Conceptualization of value is the providing of a definition, analysis or description of a value that clarifies its meaning and often its applicability" [5, p. 20]. For example, the core value of privacy might have been characterized in TM's case in terms of the control of patients' health data, data security and the possibility to remain anonymous as a patient. But with these three value qualities recognized by stakeholders, privacy is not yet sufficiently captured. From an expert perspective privacy should additionally be understood in light of the General Data Protection Regulation (GDPR) [41] or in terms of legally recognized harms [9, 42]. External sources provide a refined view on such qualities of a value that need to be considered. For the core value of privacy for instance, qualities such as data portability, data accessibility, data quality and the assurance of legitimacy of any further health data use will play a role. Note that Value-based Engineering benefits here from the clarity of its value ontology: Conceptual investigation completes and refines the qualities of core values.

**When conceptual analysis is completed, Ethical Value Quality Requirements (EVRs) must be derived for each value quality (Requirement 12).** EVRs are tangible organizational or technical requirements catering to the core value/value quality chains identified as relevant for the SOI. Taking the value quality of "informed consent" in TM's case as an example, the EVRs might include: (1) meaningful and comprehensive descriptions of personal data processing activities to TM users, (2) truly voluntary obtaining of their consent, (3) easily accessible options to decline consent and (4) measures to avoid consenting to something one does not understand. **The chain from core values to value qualities and to EVRs should be traced by organizations in what we call an "Ethical Value Register" (Recommendation 14a); i.e., with the help of a numbering system. This helps to ensure that system design will later systematically cater to values and that the ethical thought process is comprehensive.** Using such numbering systems is not only already common in risk-based approaches to system design, i.e., in security engineering [43], but it also caters to the widespread call for more transparency in system design. An illustrative continuation of the "equality" example from the TM case is shown in Figure 3, which depicts how the core value of "equality" demonstrates itself in value qualities such as "specialist accessibility" or "patient inclusion," from which five concrete EVRs are then derived (for more detail see [12]).

In the Ethical Value Register, an EVR might be described with adjectives such as "undesirable," "appropriate" or "systematic" (see Figure 3). **Minimum threshold levels or performance outcomes should be determined for an EVR (Recommendation 14b).** It is therefore recommended to add such threshold levels to the register so as to later validate whether the system lives up to the EVRs in a satisfactory way. Why such threshold determination is important becomes clear when looking at the miserable way in which many European companies today fail to even live up to the EVR of informed consent that is the bare minimum required by data protection law; many firms continue to inform their customers of personal data collection through lengthy legal texts that laypersons can hardly understand.

Many nudge users into consent, i.e., by covering half of the user screen with consent forms that do not go away unless the user agrees to personal data sharing. Unless an organization agrees to minimum EVR fulfillment standards with its stakeholders, risk remains that some will minimize later ethical design efforts as in the informed consent example.

## 4.2 Ethically Aligned Design phase

Through its process flows Value-based Engineering strives to ensure that all value qualities (and thereby core values) find entry into the design of the SOI in the order of their priority. Therefore, the Ethically Aligned Design phase ensures that the value requirements are translated one by one into concrete organizational and system requirements. Remember, system design here means "socio-technical" system design, and therefore includes not only technical requirements, but also organizational ones. People, policies and management in charge of an SOI can handle many EVRs at the organizational level without any technical dis-positions built into the system. In fact, none of the five exemplary EVRs in Figure 3 (numbered from 1.1.1. to 1.1.5) are of a technical nature. Instead they are organizational recommendations which could be immediately pursued by TM without any further technical SOI requirement definition or development needed. Consider once again TM's mission to foster more "equality" in the medical system and to help non-insured patients get treatment from recommended specialists. To fulfill this value-mission management and technology need to work hand in hand. TM needs to motivate a network of cooperating specialists who are ready to treat poor patients for an affordable price. After all, who needs a specialist database that promises "equality" if patients don't get access to the doctors recommended? The technology would be in vain.

That said, many value qualities and their respective EVRs will need to be addressed by the SOI's technical design or technology-enforced policies. Value-based Engineering has two complementary technical paths, a "low risk" design path (in the middle) is differentiated from a "high risk" design path (on the right). Why are two alternative paths necessary for EVR treatment? The reason for this is that some EVRs are derived from a core value/value quality chain of such high ethical or legal import that they require the highest possible attention and technical care. Take the core value of "privacy" in the TM case. As the company processes personal health data classified as "sensitive data" in the GDPR it has a genuine interest to not breach this law, which threatens to fine companies with up to 4% of their worldwide income if it is breached. A company like TM is therefore well advised to take the high-risk design path for the value of privacy and ensure that all EVRs are carefully addressed (for more detail see below). This advice also holds true when there is no law or international human rights agreement. **Value-based Engineering requires innovation teams to not only anticipate legally recognized value breaches, but also those that are likely to harm stakeholders' life or health. And if this is a reasonably likely scenario, then the organization should always take the "high-risk" design path, which is risk-assessment based. Otherwise the alternative "low-risk" design path can be chosen. (Requirement 13).**

### 4.2.1 Design path for low-risk EVRs

The system engineering activities in the low-risk design path should in many respects be compatible with those prevalent in today's iterative and agile forms of system design and development. In principle, companies could take the low-risk EVRs and simply employ them with their existing design and development processes. Still, we want to point to four activities particularly recommended for Value-based Engineering: (1) making use of a broad indirect persona analysis; (2) deriving EVR architecture and design treatments with the help of a (light) risk logic; (3) creating a holistic prototype which integrates the EVRs with other functional system requirements stemming from non-ethical requirements engineering; (4) working with the notion of a "first viable product" as is done in Design Thinking [17].

When it comes to persona analysis, then this is a well-known usability practice [18, p.221f]. It uses representational figures that act as stand-ins for real users of an SOI. **For Value-based Engineering we recommend that the type of persona included in the system design analysis should be complemented by indirect stakeholders (Recommendation 15).** In TM's case, for instance, the TM network of doctors who fill the database with recommended specialists should be represented by a persona. What kind of interface would they need to ensure their judgments can be trusted? Would they want to stay anonymous? Or, would it in contrast be essential for TM to be transparent about the "judges" filling their database? How can fairness and professionalism be insured in the way recommendations are made? These questions show that when trust, reliability, accuracy and honesty are highly prioritized core values for TM, then their indirect persona inclusion in the design process is essential. Indirect persona inclusion ensures that a broader group of stakeholders are considered not only in the higher-level exploration of values, but also in the concrete application of these in the system design.

The second recommendation for low-risk Value-based Engineering is that the iterative design of an SOI should follow a risk logic (Recommendation 19). Risk-driven design approaches are not new to engineering. What risk-driven design generally does is that it asks for each SOI requirement whether it is at risk of not being fulfilled, and what the "threats" are that may cause this non-fulfillment. Technical design then seeks to address the respective threats and mitigate them one by one through corresponding controls we call "value-risk-control requirements" here. Generally, a threat/control analysis for value-based system design can be very lengthy and detailed as for instance the NIST standard for system security [9]. Value-based Engineering doesn't recommend their use in fullest depth for low-risk EVRs, but only for high-risk EVRs (see Path 3 described below). But even for the low-risk EVRs we believe in line with other computer ethics scholars [44] that conceiving of values as "being-at-risk" puts development teams into a spirit of care and awareness for what could go wrong. **Companies engaging in Value-based Engineering should therefore explore how they could accommodate such "risk-thinking" in their established design and development processes (Recommendation 16).** Thirdly, value-risk-control requirements should find an entry into SOI mock-ups or prototypes, which are then iteratively tested with real internal and external users. Note, though, that such prototypes cannot be exclusively based on the value analysis presented in this paper. Instead, the identified value-risk-control requirements need to be integrated with the other functional system requirements that do not stem from the ethical thinking process. Organizations have functional systems in place and might just want to improve them with Value-based Engineering. Or they have developed functional requirements to make the SOI work technically. **Value-based Engineering requires that the technical system requirements from the functional engineering domain are integrated with the value-risk-control requirements (from the ethical domain) into one "holistic system design concept" (Requirement 14).** This holistic system design concept embedded in system mock-ups, prototypes or new design versions of an existing system should live up to the EVR threshold standards described above. Figure 4 summarizes the conceptual integration of the two domains—ethical and functional—in one holistic system design concept.

Finally, the most promising holistic system concept will enter a test market for real-world feedback. **It is recommended that continuous feedback is collected from stakeholders on the first and later viable SOI versions (Recommendation 17).** This feedback is not only focused on the usability or user experience of the SOI, as most classical system design is; rather, Value-based Engineering looks for complementing value qualities, EVRs or even core values that may have been missed in the value exploration phase. It also observes the reaction of the wider set of indirect stakeholders rather than direct SOI users only. Various stakeholders might put the SOI to unethical uses not anticipated. Equally, the SOI might produce unexpected negative values, such as addiction, hate or long-term unease; that is, negative values that were not expected can appear. Such values would then be re-inserted either into the value prioritization activity described above, or be added to the EVR identification process. Additionally, it might turn out that protection against anticipated negative values is not as successful as planned. Value-based Engineering includes the monitoring of how the treatments chosen are effective in mitigating value threats later. All in all, the theoretical value exploration phase now becomes a life-exploration phase that iteratively circles with the design process to continuously improve the SOI.

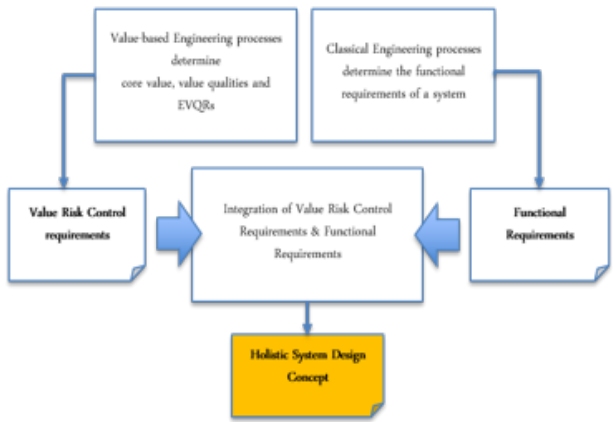

Fig. 4. Conceptual integration of ethical and functional concepts

### 4.2.2 Design path for high-risk EVRs: risk assessment-based design

While the immediate organizational and low-risk path will be sufficient to address the majority of EVRs, some value qualities are so vital for stakeholders that system design needs to be even more rigorous. In such cases it is more difficult to integrate already existing or external system elements or functional system requirements that have not been developed with a risk logic. It is necessary that organizations build their SOI from scratch and potentially also migrate existing systems into the new risk-aware infrastructure.

Risk assessment-based design methods are well established for some values, such as security, privacy and safety [9]. However, there has not yet been an attempt to generalize these methods to a generic value level. This is necessary for Value-based Engineering. **When Value-based Engineering necessitates taking risk-assessment-based-design as a guiding template [9], then design starts logically at the core value/value quality level (Recommendation 18).** For instance, the core value of TM's patient privacy is given through the value quality of health data security. This value quality of security can be met provided that the ethical value quality requirements (EVRs) of confidentiality, integrity and authenticity are well respected in the design of TM's SOI. All of these EVRs could be the target of an attack. So the first step in risk-assessment-based design is to understand the level of protection demand required for them. The level of protection demand is determined by asking, "What would happen if for various stakeholders the EVR was not met?" For instance, what would happen if the confidentiality of TM patients' health data was breached? This question can help to understand and weigh the EVR's level of protection demand, normally on a nominal scale. Afterwards, each EVR undergoes a similar threat analysis as outlined above. While in security and safety assessments such a threat analysis can comprise a relatively objective calculation of threat probability, many human or social value threats might not be that easily quantifiable. Therefore, it suffices to judge whether a threat to an EVR is realistic or not. And for those that are, each one needs to be addressed and mitigated by a respective system control. System controls can come in functional, non-functional, operational, procedural, organizational or structural form. However, there are typically alternative control options feasible with different degrees of rigor. The choice of the appropriate control option should be informed by the level of protection demand identified for the EVR.

Finally, it is recommended that the whole process of deriving control requirements is documented and traceably linked back to the respective value qualities. The effectiveness of the controls chosen and implemented as "value dispositions" in the SOI (see 3.4) should be monitored during market deployment and adapted throughout the life cycle of the system. This can be done in the same way as described above for low-risk design. The iteration(s) might be done by that party which continues to service the system.

### 4.3 Transparency management for ethics by design

Genuinely intending and creating value dispositions in products has been described above as a core ambition of Value-based Engineering. The virtue that lies in this form of activity needs to be appreciated in itself. Therefore, the artifacts produced in each process described above should be documented in the "Ethical Value Register." **The Ethical Value Register we recommend to build up can serve project management and later auditors to recap at any time what the goals of the project were, who was involved and who signed off on it (Recommendation 19).** It should contain the full spectrum of conceptually analyzed core values and value qualities depicted in their clusters as well as the priorities agreed on. The stakeholders involved should also be noted as well as their agreements and disagreements on individual value qualities and EVRs. The names and personal signatures of those top executives who endorse the priorities should be captured (see recommendation 11). Furthermore, it is recommended that the value mission statement described above is made publicly available (see recommendation 15). Finally, engineers have to back the degree of risk they are willing to take when risk-assessment-based design choices are made. Subsequently, **when they decide on a level of control to treat a value threat, then such choices should be recorded in the Ethical Value Register and endorsed by the respective engineers who are responsible for it (Recommendation 20).**

The documentation effort has of course the disadvantage of costing extra time and effort that is not en vogue in times of low-cost and agile system development. However, there are two key advantages to documenting these information items that go beyond mere validation. First, it is possible to develop a culture of responsibility and transparency that is adept at tackling things that go wrong. Take the case of Volkswagen, where the culture was not as open, and where it took months to understand who took the decision, when, and in what context to build misleading software for the cars' emission statistics. Such lengthy searches for culprits and scapegoats traumatize engineering departments and all those involved. Secondly, one should not underestimate the power of the performative act that is putting one's good name on the line. The physical act of standing by something in written form is likely to foster accountable behavior, to support more appropriate risk taking and to increase customer trust. When no one wants to put down his or her name for a system design choice or value priority, then this is also a good indicator during engineering for rethinking a respective decision.

## 5 Challenges for Value-based Engineering

Value-based Engineering is a new way to approach technical innovation: A "value-mission" is built from the human, social, environmental and technical value priorities upwards and with the help of an ethically driven stakeholder discourse. Stakeholders include not only direct end users, customers or developers as prime engineering standards have stressed so far [25], but also indirect stakeholders, nature and society at large. Their voices determine the value principles that are then systematically translated into system practice. Corporate leaders' role in this is not simply to maximize profit anymore. Instead their role is to help prioritize the value space uncovered in an ethical way. Thereby financial value becomes only one indicator of corporate success among other values. With this sweeping vision, Value-based Engineering is not easily implementable in today's business context. Instead it will face a number of challenges grand and small.

### 5.1 A change of work culture

Value-based Engineering requires an evolution of software engineering culture towards ethical diligence. While agile forms of software engineering have been heralded in recent years as a breaking free from rigid and reporting-loaded work [45] leading to rapid results [46], Value-based Engineering requires some backpedaling. Time must be taken to think about value requirements and risk management, and also to write the necessary documentation. This time often goes beyond the scope of tight budget plans, and might be seen as undermining profit as development costs are likely to rise. In this environment it is not easy to embrace the degree of ethical responsibility that goes hand in hand with the Value-based Engineering processes described here. At the time of writing, many technical engineers have gotten used to delegating ethical responsibility to legal departments [47, 48]. It has been reported that 40% of them do not even feel responsible for the privacy and security of their systems [49]. Handling ethical principles is often regarded more as a compliance necessity than as a positive challenge.

Against the background of this reality, Value-based Engineering can imply a change of work culture both at the managerial and technical level. Engineers might be empowered to embrace to a much greater extent the codes of ethics that professional associations like the ACM have formulated for their members [50]. A true challenge here will be that technical engineers might need to work with business managers more intensely, because so many EVRs require organizational action or policies that are then embedded in the SOI. Such cooperation between engineers and managers has proven difficult in the past, because management and engineering departments have very different work cultures [18].

### 5.2 New roles and training curricula

Engineers and managers with their given education and experience will have difficulty embracing Value-based Engineering without support. As was shown above, the value exploration phase needs a new kind of employee: a value expert. People in this role need to have a faculty of judgment that can only be obtained from interdisciplinary education or coaching that is still rare today: knowing the humanities with a focus on ethics and values, and having a good understanding of technology and management as well.

Ideally, value experts should not be independent consultants who leave a project once a first system version is deployed. Since there is iteration and monitoring foreseen as part of risk management, a person permanently embedded in the development team should take up this role—for example, someone who has the longer-term "power user," "scrum master" or "system engineering" role in development efforts, or who is the permanent product manager of the system.

Investing into this kind of ongoing new role(s) in organizations will need to be shouldered financially, and new educational programs are needed to educate the experts filling this role. Moreover, value experts need to work with stakeholder representatives who must first be identified and then convinced to cooperate, perhaps in exchange for some financial compensation. While it is true of course that companies have been working with stakeholders already and succeeded in engaging them, the kind of stakeholder representative foreseen by Value-based Engineering is different. The scope of potential stakeholders is widening by reaching out to NGOs and people with a critical perspective. Such stakeholder representation might not be so easily attracted and maintained.

### 5.3 More need for control, accountability and privacy

Value-based Engineering seeks a high degree of control over engineering artifacts: the ecosystem of partners therefore needs to be accessible and manageable. This culture of control-based responsibility is challenged by the way software is built today. Smaller companies in particular do not always challenge the details of the ecosystem they depend on. They tend to copy requirements from established applications [1], leading to a progression of biases [51] or other unethical practices. Furthermore, today's software is often built out of pre-developed functional modules. Less than 30% is custom-built [34]. Such modules, however, are often proprietary, not allowing in-depth investigation. If access is given, engineers might lack the special knowledge or time necessary to fully understand them. A related challenge is that the complexity of module integration can cause unpredictable interactions. This potential for unpredictable interactions makes reliable upfront planning (as suggested for Value-based Engineering) a challenge [34]. Take again the case of TM that only has a five-person budget and needs to integrate third-party video conferencing software in its service if it ever wants to go live. In this situation, it might not be able to control the security level and data handling policy of its video software partner.

Even if TM finds an appropriate partner, the next question is on what terms and conditions cooperation is offered. Often personal data or access to customers is part of the service agreements, even though this is recognized as a privacy breach [1]. In today's data-driven digital economy, the company is still forced to toy with the idea of selling its patient diagnosis data at some point, considering it as a valuable source of profit. The ethical necessity to forgo data sharing agreements is not easy for such a small company. It would then depend solely on its own service quality or need to engage in higher sourcing costs for services, which may be the right way to go from an ethical perspective, but not necessarily the one desired by many of today's investors.

### 5.4 Readiness for one's own ethical truth

Value-based Engineering demands a high level of personal maturity from everyone involved in it. As the authors learned from the TM case study, being confronted with the negative values one might create through one's SOI or business model is psychologically difficult to swallow for managers who want to succeed with their idea. We "tend to adjust our value judgments to our factual willing and acting (and our weaknesses, deficiencies, faults, etc.)," wrote Scheler about the challenge of truly facing the values one creates [24, p. 327]. In TM's case, the CEO needed to face the fact that he might be breeding distrust, envy and competition among doctors if his service ever reached a relevant market share. The message that the price of TM might be to gain a culture of efficiency in exchange for a culture of benevolence is not a message that was easy to swallow. Value-based Engineering therefore seems to be advisable only in an early innovation phase where there is openness towards the formation of a value mission. Or, where a company is in a situation of such ethical turmoil that it is ready to radically turn around its business with Value-based Engineering, regardless of the changes and costs to the business model and technical infrastructure. In TM's case the core value briefing and feedback on ethical challenges came one year into the project and the CEO did not want to hear that doctor competitiveness and distrust could undermine his business idea. In the end he lost an additional year on talks with doctors' associations which eventually declined cooperation with TM, so that ultimately he abandoned the idea of ranking altogether.

### 5.5 The challenge of measuring values and ethical conduct

Finally, it is a fair question to ask how to measure the success of a Value-based Engineering effort. How can the "success" of such an effort be judged if so many people and cultures tend to hold different values? We argued that engineers and value experts need to trust their "value feelings." But is this rational, objective, reliable? Is it not objectively impossible to measure the degree of value attainment? The malleability of values is indeed a challenge for those who prefer to trust only that which can be precisely defined and linear-causally determined [52]. Albrechtslund has called this challenge "the positivist problem" [53]. Since not all values are definitely measurable, it is hard to prove at the end of a project that the value mission a system set out to create has actually been achieved. While User A might appreciate the control, transparency or privacy a system grants, User B might not even recognize any of these values as being present. Depending on our individual training, experience, preferences, etc. (or as Scheler calls it "milieu" [24]), the perception of values varies greatly from one human to the next. Therefore, the proof of a technical product being "ethical" can hardly be based on the purely quantitative value judgments people make once a product is launched. Therefore, we recommend that narratives and experiences should be used to gain insight into the success or failures of Value-based Engineering.

That said, there is a way to demonstrate that a system is ethical. This is due to the "act related value" [15, p.252] inherent in engineers' objectively verifiable effort to create an ethical product by following the processes described. One can observe how many core value clusters are addressed in system design and based on this, an "ethical maturity" level can be measured. After all, engineers deserve to be called "ethical" if they regard their systems as "value bearers" [15, p. 17, 122, 105] and demonstrate their effort to create positive value qualities by embedding the respective "value dispositions" into the technology. Noting that there are also negative values, engineers equally deserve to be called "ethical" if they genuinely intend to avoid the nega-tive value qualities undermining core values (i.e., uncontrollability, opacity, insecurity) and take the necessary precautions. In other words, if engineers care, their work itself becomes an ethical activity which is guided by the honest desire to do good. The documentation in the Ethical Value Register is the objective proof of that intent. Therefore, Value-based Engineering is transparent not to prove compliance, but in order to show how the technology was created for humanity.

## 6 CONCLUSION

The Value-based Engineering methodology presented here is a new vision on how to innovate for a better world, in which it is no longer corporate profit that primarily determines technology development, but the motivation to innovate for humanity. Deeply rooted in material value-ethics, moral philosophy, discourse ethics, VSD as well as system and software engineering, this article summarizes how Value-based Engineering works; that is, what processes it needs to be lived in practice. More precisely 14 process requirements and 20 recommendations are elaborated, which organizations can embrace. In addition, new terms are introduced to describe value dynamics and make these accessible for engineering; in particular, we outline how a value mission can be ethically formed around core values. Core values and their qualities are prioritized and translated into technical and organizational practice. A new entity we call "Ethical Value Quality Requirement," or "EVR" is introduced to make the (long-missing) link between relevant stakeholder values and a technical design effort. This effort needs to be guided by a new kind of employee whose role we propagate here—that of the ""value expert." A value expert is a shepherd understanding both the technical and ethical world and helps to guide SOI development over time. Whether such an endeavor to propose a new form of ethical system design can become successful is unknown. For the most part the method described here with all its activities, terms and roles are in line with and in response to the work of experts who have been involved in IEEE's P7000 standardization project. P7000 was set up by IEEE in 2016 to establish a process model which addresses ethical considerations throughout the various stages of system initiation, analysis and design. Thousands of comments from workgroup members and standard balloters have informed a majority of the views outlined in this paper. History will show whether the ideas bear value.

## ACKNOWLEDGMENT

The authors are grateful for interaction with participants of the IEEE P7000 working group, which held more than 100 calls over the span of almost five years to practically discuss the matters described in this article. We express our exceptional gratitude to Ali Hessami, Lewis Gray, Ruth Lewis, Rob Schaaf, Annette Reilly and Gisele Waters as well as many other active supporters of this effort. We want to thank the CEO of TM for the opportunity to do the case study. Finally, we are grateful to our colleague Kathrin Bednar for supporting us in the academic analysis of the TM case study.

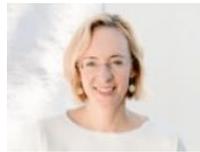

**Sarah Spiekermann** is a Ph.D (2001) and tenured professor for Information Systems at Vienna University of Economics and Business where she chairs the Institute for IS & Society (since 2009). She has authored over a hundred scientific articles, including publications in IEEE Proceedings, IEEE Transactions on Software Engineering, IEEE Security and Privacy, Communications of the ACM, Journal of Information Technology (JIT), and many other leading outlets. She was co-editor of the European Journal of Information Systems (EJIS), author of the textbook "Ethical IT Innovation – A Value-based System Design Approach" as well as the books "Networks of Control" on personal data markets (2016) and "Digital Ethics – A Value System for the 21st Century" (2019). Sarah has been a member of IEEE since 2016 and has been vice-chair of the P7000 effort.

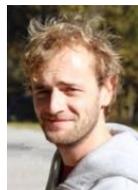

**Till Winkler** is a Ph.D. student since 2017 at the Institute for IS & Society at the Vienna University of Economics and Business. He has been a workgroup member of the IEEE P7000 effort as a volunteer since 2017.